\documentclass[]{spie}  %>>> use for US letter paper
\usepackage{amsmath,amsfonts,amssymb}
\usepackage{graphicx}
\usepackage[colorlinks=true, allcolors=blue]{hyperref}

\title{The DESI Sky Continuum Monitor System}

\author[a]{Suk Sien Tie}
\author[b]{David Kirkby}
\author[a,d]{Paul Martini}
\author[c]{Claire Poppett}
\author[a]{Daniel Pappalardo}
\author[c]{David Schlegel}
\author[a]{Jonathan Shover}
\author[c]{Julien Guy}
\author[d]{Kevin Fanning}
\author[d]{Klaus Honscheid}
\author[c]{Michael Lampton}
\author[c]{Patrick Jelinsky}
\author[c]{Robert Besuner}
\author[c]{Kai Zhang}
\author[c]{David Brooks}
\author[c]{Peter Doel}
\author[c]{Yutong Duan}
\author[c]{Enrique Gasta$\tilde{\mathrm{n}}$aga}
\author[c]{Robert Kehoe}
\author[c]{Martin Landriau}
\author[c]{Michael Levi}
\author[c]{Francisco Prada}
\author[c]{Gregory Tarle}

\affil[a]{Department of Astronomy, The Ohio State University, Columbus, Ohio, USA}
\affil[b]{Department of Physics and Astronomy, University of California, Irvine, California, USA}
\affil[c]{Lawrence Berkeley National Laboratory, Berkeley, California, USA}
\affil[d]{Center for Cosmology and AstroParticle Physics, The Ohio State University, Columbus, Ohio, USA}

\authorinfo{Further author information: (Send correspondence to S.S.T.)\\S.S.T.: E-mail: tie.5@osu.edu}

\begin{document} 
\maketitle

\begin{abstract}
The Dark Energy Spectroscopic Instrument (DESI) is an ongoing spectroscopic survey to measure the dark energy equation of state to unprecedented precision. We describe the DESI Sky Continuum Monitor System, which tracks the night sky brightness as part of a system that dynamically adjusts the spectroscopic exposure time to produce more uniform data quality and to maximize observing efficiency. The DESI dynamic exposure time calculator (ETC) will combine sky brightness measurements from the Sky Monitor with data from the guider system to calculate the exposure time to achieve uniform signal-to-noise ratio (SNR) in the spectra under various observing conditions. The DESI design includes 20 sky fibers, and these are split between two identical Sky Monitor units to provide redundancy. Each Sky Monitor unit uses an SBIG STXL-6303e CCD camera and supports an eight-position filter wheel. Both units have been completed and delivered to the Mayall Telescope at the Kitt Peak National Observatory. Commissioning results show that the Sky Monitor delivers the required performance necessary for the ETC.
\end{abstract}


%\noindent
%\textcolor{red}{Note: I don't hyphenate `back illumination' as a noun or verb, but only when it acts as an adjective before a noun, e.g. `back-illuminated' fibers.}

\section{INTRODUCTION}
\label{sec:intro} 
%1) DESI background \\
%2) Dynamic ETC introduction; estimated gain from dynamic vs fixed exposure time. \\
%3) rationale for Sky Monitor existence \\

The Dark Energy Spectroscopic Instrument (DESI) is a Stage-IV experiment by the US Department of Energy to measure the dark energy equation of state with high precision\cite{Martini2018}. DESI will achieve this by spectroscopically mapping 34 million galaxies and quasars, an order of magnitude increase compared to previous surveys, over a five-year period. 

Previous major spectroscopic surveys like the Sloan Digital Sky Survey (SDSS) and BOSS (Baryonic Oscillation Spectroscopic Survey) adopted a fixed exposure time regardless of observing conditions. To ensure a uniform SNR depth of the data, these surveys relied on real time reductions of actual spectra to estimate the SNR of each exposure, after which decisions were made whether to continue observing with additional exposures. DESI will not adopt this approach as most DESI targets typically require only one or two exposures. Instead, it will utilize real time observing conditions to estimate the optimal exposure time that will meet the SNR requirement for a fiducial target. Besides maximizing survey efficiency, this also ensures a uniform depth in the spectroscopic data and more uniform redshift completeness, which are crucial for cosmological surveys. 

The dynamic exposure time calculator (ETC) is responsible for adjusting the spectroscopic exposure time during each science exposure. It tracks the SNR/pixel for a fiducial source and closes the spectrograph shutter once a desired integrated SNR has been achieved. The ETC relies on the DESI guide cameras \cite{Jimenez2016} for the signal estimate, particularly to extract the atmospheric seeing and sky transparency. In principle, the noise estimate, i.e. the sky level, can be estimated from starless regions of the guide cameras. However, this is not expected to be sufficient as the sky level is degenerate with the dark current, which varies enough with temperature that it may not be possible to measure the sky brightness with enough accuracy. Alternatively, starless regions of the \textit{focus} cameras can be used instead, as each has a dark occulting bar that allows separation of the dark current and sky brightness. Here we present an alternative solution: a dedicated imaging system -- the Sky Monitor -- that will be used to monitor the sky flux down the DESI sky fibers. Information from both the guide cameras and the Sky Monitor will be continuously fed to the ETC over the course of a science exposure. 

In this manuscript, we describe the design of the Sky Monitor in Section \ref{sec:design} and the commissioning and operation of the Sky Monitor in Section \ref{sec:operation}. We summarize the main results and lessons learned in Section \ref{sec:conclusion}.

\section{DESIGN}
\label{sec:design}

\subsection{Mechanical Design}
\begin{figure*}[h]
\centering
\includegraphics[width=0.6\textwidth]{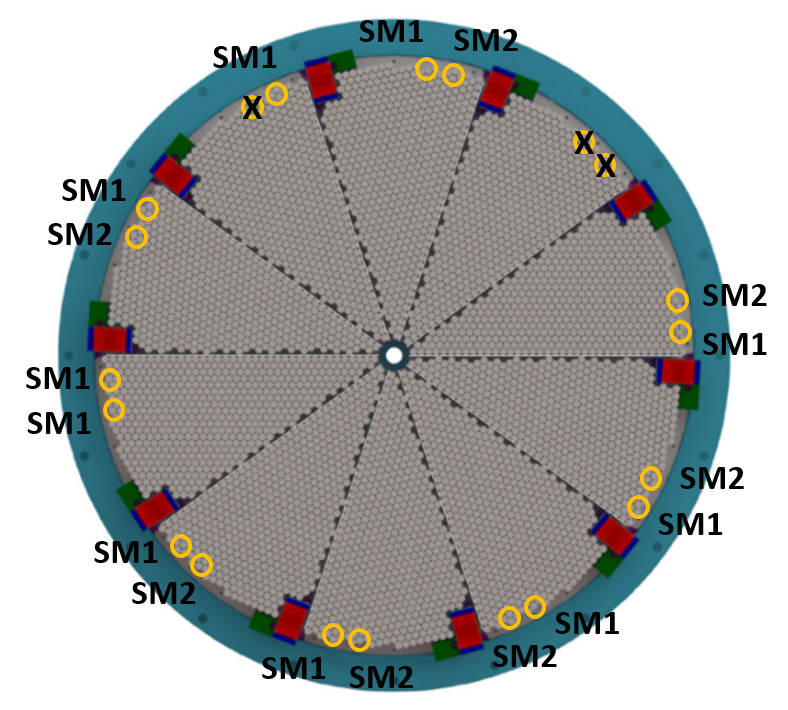}
\caption{\small The DESI focal plane with the sky fibers marked in yellow circles. The DESI focal plane is divided into ten pie-like modules, or ``petals''. With two sky fibers per petal, there are a total of twenty sky fibers, of which seventeen are operational. A black cross denotes a broken sky fiber. Ten fibers are currently fed into Sky Monitor 1 (``SM1'') and seven into Sky Monitor 2 (``SM2''). The red rectangle on each petal denotes either the focus camera or the guide camera.}
\label{fig:layout}
\end{figure*}

 The Sky Monitor will measure the sky background level via seventeen sky fibers located on the periphery of the DESI focal plane. These sky fibers are split between two identical units to provide redundancy in the event of a failure of one unit (the ETC can continue operating with only a subset of the sky fibers). Figure \ref{fig:layout} shows the locations of the sky fibers on the DESI focal plane and which Sky Monitor unit they are connected to. The DESI sky fibers from the focal plane terminate at the spectrograph fiber spool boxes. Fiber spool boxes are used for fiber routing management and are located at each end of the fiber cable. There are ten spool boxes connecting ten petals to ten spectrographs. There are two sky fibers from each petal to each spool box. Rather than being subsequently routed to the spectrographs like the science fibers, the sky fibers are routed to the Sky Monitors. We used commercial patch cables constructed with DESI fibers and SMA905 connectors to connect the spool boxes to the Sky Monitor. The fiber tips of these patch cables have been AR-coated. 

\begin{figure*}[h]
    \begin{minipage}[b]{0.50\linewidth}
        \centering
        \includegraphics[width=1.2\textwidth, angle=270]{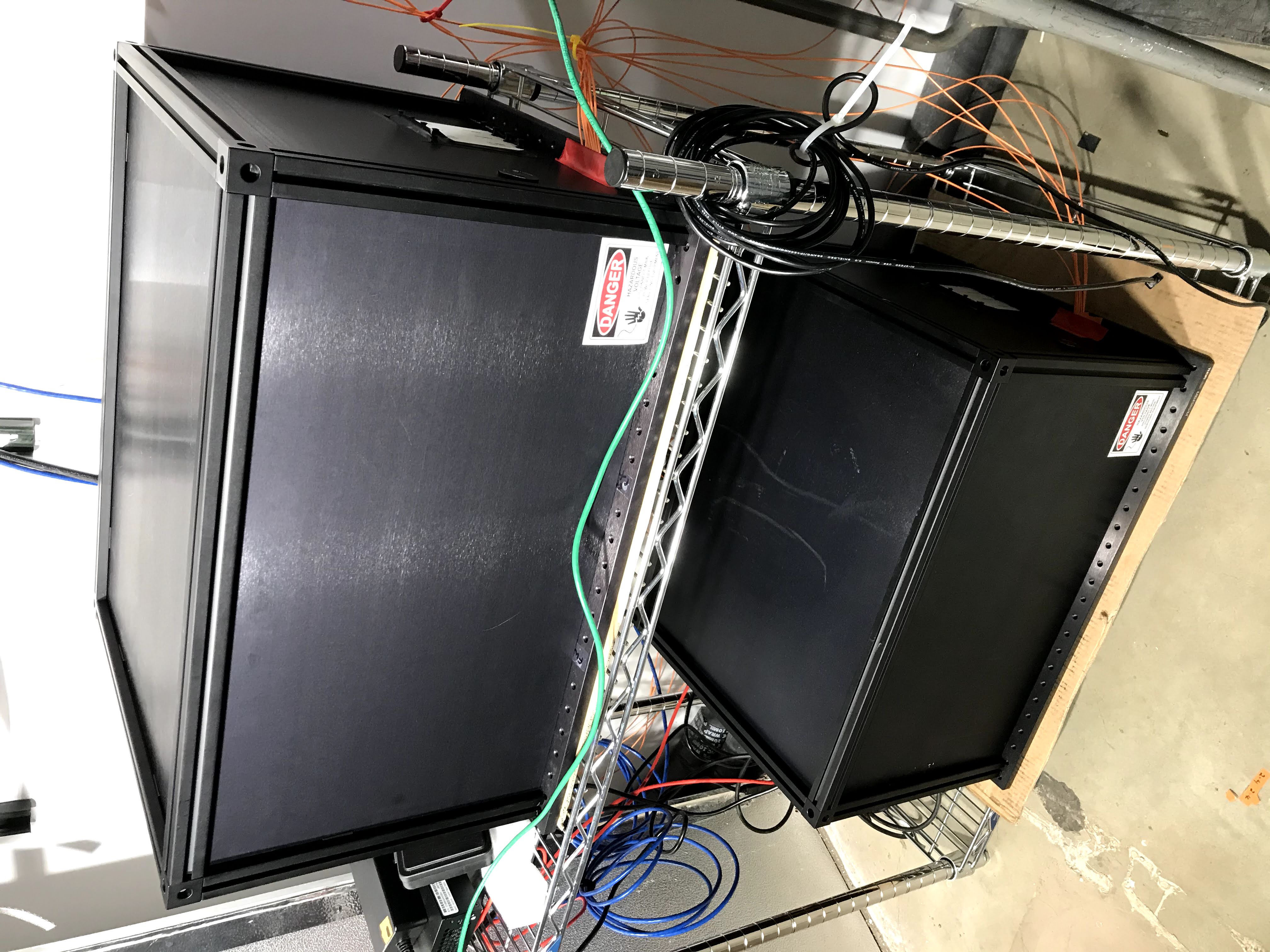}
    \end{minipage}
    \hspace{-8mm}
    \begin{minipage}[b]{0.50\linewidth}
        \centering
        \includegraphics[width=1.2\textwidth, angle=270]{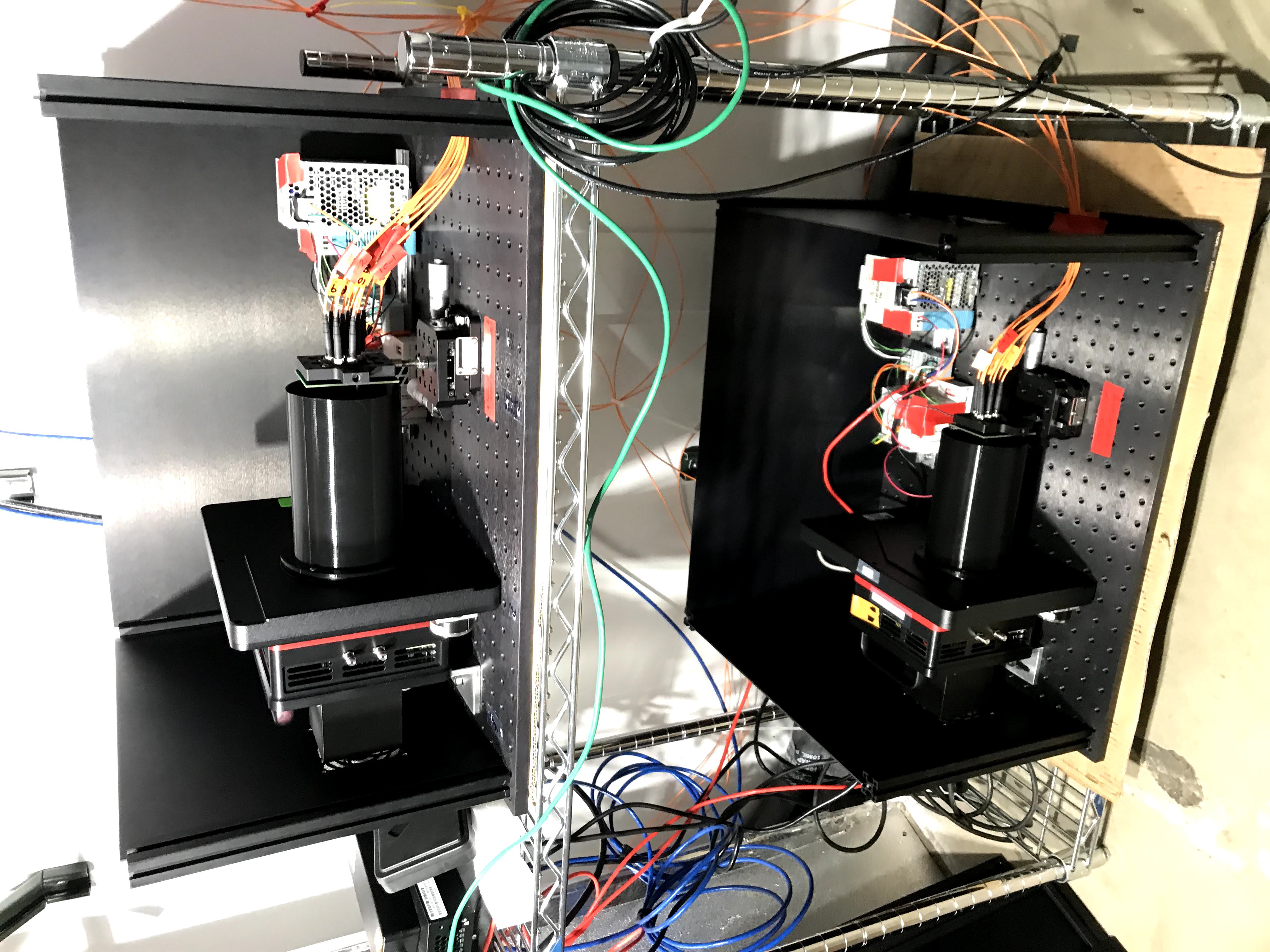}
    \end{minipage}
\vspace{2mm}
\caption{\small The two Sky Monitors, fully enclosed (left) and partially opened up (right).}
\label{fig:external}
\end{figure*}

Each Sky Monitor weighs $\sim$ 21 kg and is $\sim$ 525 mm $\times$ 375 mm $\times$ 300 mm in size. Figure \ref{fig:external} shows the Sky Monitors fully enclosed (left) and with the top and front enclosures removed (right). Each unit is mounted on a breadboard and enclosed in commercial black aluminium enclosures to minimize light leakage. We cut mouse holes on the left and right side panels for the sky fibers and the power cables, respectively. We also created an opening on the left side panel to provide an entry point for the camera fan air flow, which is covered with a dust foam cover.

\begin{figure*}[h]
\centering
\includegraphics[width=1.0\textwidth]{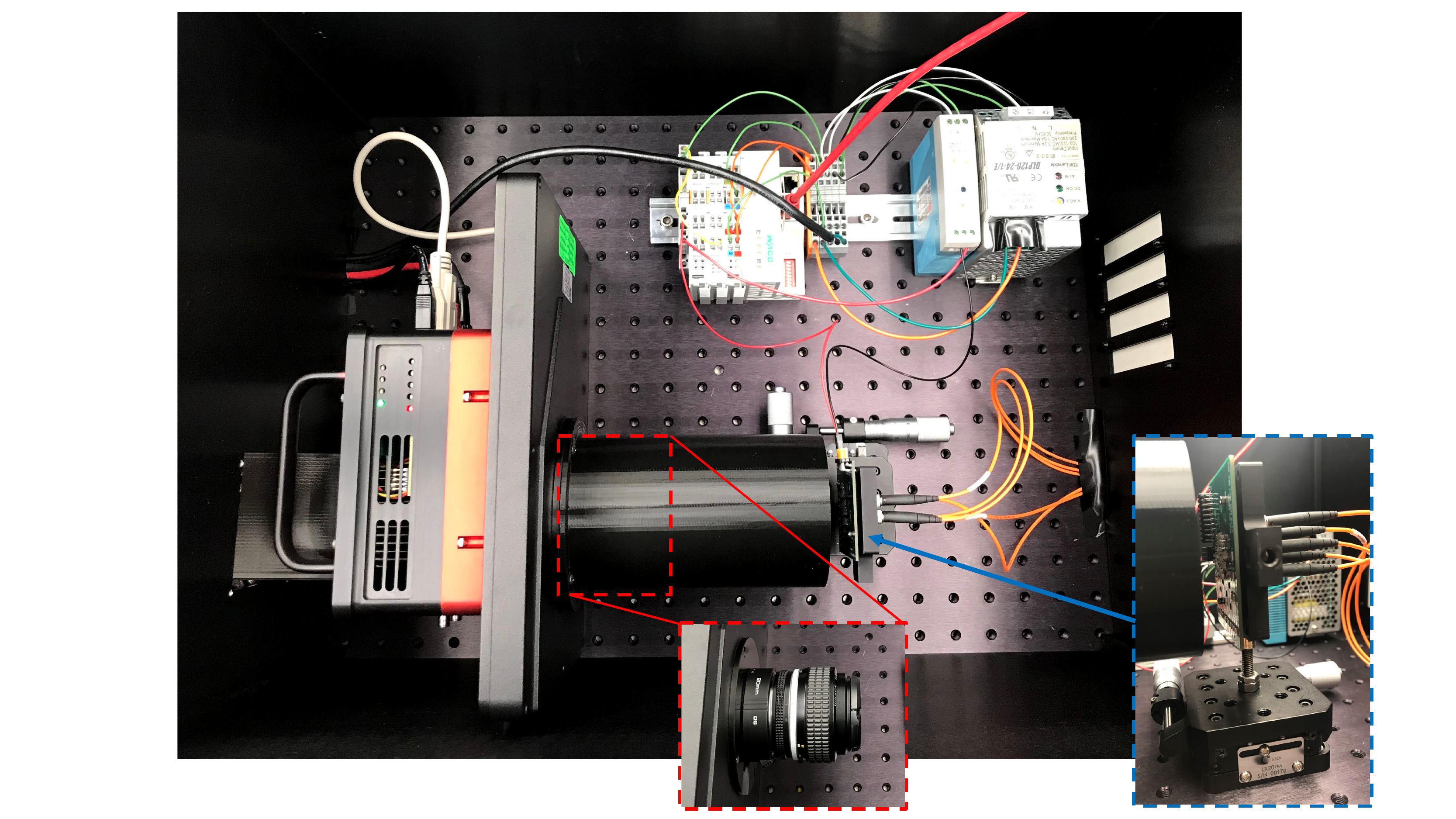}
\caption{\small Internal view of the Sky Monitor, consisting of a CCD camera, a filter wheel, a lens that is enclosed in a 3D-printed baffle tube (red box), mounting system for the sky fibers (blue box), and electronics for the back illumination system. The mounting system is made up of a custom fiber block that holds the sky fiber, which is mounted on an XY translation stage.}
\label{fig:internal}
\end{figure*}

% imaging component
The Sky Monitor is made up of three main hardware components: an imaging camera, a mount for the sky fibers, and a back illumination system. Figure \ref{fig:internal} shows a top-down view of the system. The imaging components consist of a CCD camera, an eight-position filter wheel, a commercial lens, and custom 3D-printed baffle tube and air vent. The Sky Monitor reuses the SBIG STXL-6303e CCD cameras that were used for the DESI Commissioning Instrument\cite{Ross2018}. Their fast readout rate (measured full frame download time of 5.5 seconds with its linux driver) and support for binned readout mode also make this camera a suitable choice. The SBIG camera mates with the eight-position FW8S-STXL filter wheel, which supports 50 mm diameter 3 mm thick filters. The filter wheel takes $\sim$ 2.5 seconds to move to the next slot and only rotates in one direction. 

The filter wheel is currently populated with an SDSS r-band filter (562 $-$ 695 nm) and a filter stack for the back illumination (more details below). An r-band filter was chosen as it covers the wavelength range crucial for the redshift determination of emission-line galaxies targeted by DESI, whereas the broadband SDSS filter was used to improve the faint sky flux signal. An appropriate lens is needed to produce an overall compact system and minimize lens distortion without vignetting the sky fibers. Our selection criteria led us to the Nikon 50 mm f/1.2 lens, which was coupled to a 20 mm extension tube to achieve a magnification of 0.5. A custom 3D-printed black baffle tube is attached around the lens to minimize stray light. Finally, a 3D-printed air vent is also attached to the camera fan to direct air flow.

% mounting system
The sky fibers are held in a custom fiber block assembly, as shown in Figure \ref{fig:internal}, which is then attached to a commercial manual XY translation stage; this forms the mounting system. The translation stage is positioned such that we can move the fiber block in the focus and lateral directions. 

% back illumination
In addition to collecting light down the sky fibers, the Sky Monitor also back illuminates these fibers to allow repositioning in the event that a star lands on a sky fiber. The DESI fiber assignment code allows us to know in advance if a fiber will fall on a star, and if so, to reposition it as needed. Fiber positioning requires precise knowledge of the current location of the fibers, which is achieved by back illuminating the fibers and imaging them with the Fiber View Camera (FVC)\cite{Baltay2019} that is located behind the central hole of the primary mirror of the Mayall telescope. The Sky Monitor back illumination system consists of a custom printed circuit board that holds eighteen 460 nm (the wavelength that the FVC detector is most sensitive to) surface mount LEDs, LED control electronics and power supply, and a diffuser/neutral density filter stack. We ran lab tests to determine the number of LEDs needed to be comparable to the brightness of the exposure shutter illuminator\cite{Derwent2016} that back illuminates the fiber positioners.

The LED circuit board is mounted in front of the fiber block such that the LEDs face the filter wheel. The filter stack, made up of a ground-glass diffuser and a 99.9\% reflective metallic neutral density filter, is configured such that the LED light first hits the diffuser before being reflected by the neutral density filter. As such, the filter stack reflects diffuse light into the sky fibers so that the fiber tips can be uniformly illuminated. Additionally, the LEDs are tightly packed within an allowed region on the circuit board so that their reflected light uniformly illuminates the acceptance angle of the fibers. %The LEDs emit a constant brightness that is comparable to the brightness of the exposure shutter illuminator\cite{Derwent2016} that back illuminates the fiber positioners. 

\subsection{Electrical Design}
The electrical components of the Sky Monitor consist of the CCD camera, the WAGO ethernet fieldbus controller (as part of the back illumination control electronics), and the LED circuit board. These components require electrical power and rely on ethernet connection for data transfer and communication. The power and ethernet connection to both Sky Monitor units are provided by a Raritan power distribution unit and an ethernet switch.
%\textcolor{red}{The eighteen LEDs on the circuit board flash simultaneously at a cadence of (???) seconds to produce the appropriate brightness level.}

From the Raritan power distribution unit, one power cable serves the CCD camera and another serves the 24 VDC power supply for the back illumination control electronics. The back illumination control electronics subsequently powers an internal 5V power supply for the LEDs. The CCD camera and the back illumination control electronics are controlled by the DESI Instrument Control System (ICS)\cite{Honscheid2018} via an ethernet connection. From the ethernet switch, an ethernet cable connects to each of these components. 

\subsection{Software Design}
The Sky Monitor operations comprise taking exposures, changing the filter wheel, and turning on/off the back illumination LEDs. These operations are done through the ETC software and integrated with the DESI ICS exposure control. 

\begin{figure*}[h]
\centering
\includegraphics[width=0.8\textwidth]{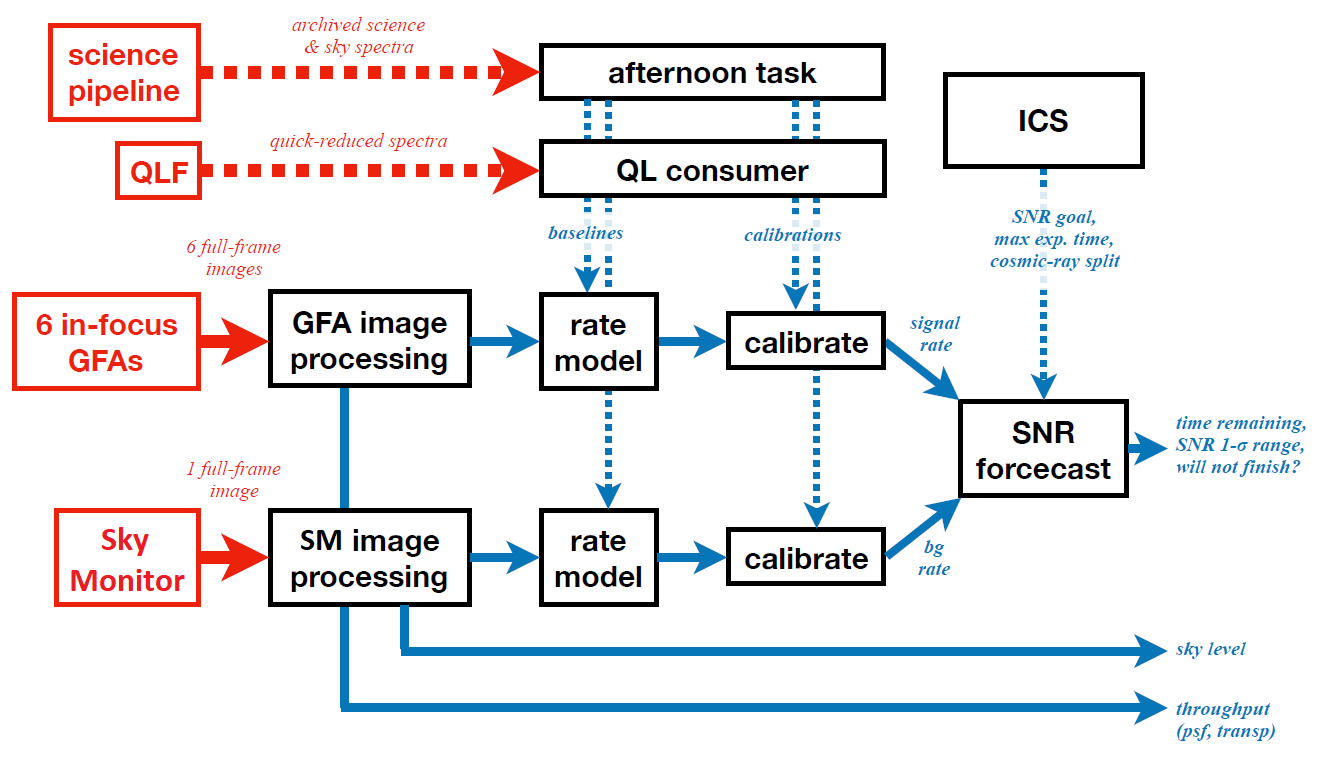}
\caption{\small Input data and processing flow of the dynamic Exposure Time Calculator (ETC) in producing a real-time signal-to-noise ratio (SNR) forecast. The red boxes denote input sources and the black boxes denote the analyses of these input data. The solid lines refer to the base processing loop that occurs nightly and the dashed lines refer to day-time or additional night-time analyses. The guide cameras (`GFAs') and the Sky Monitor are mandatory components of the dynamic ETC. The output of the Sky Monitor (guide cameras) serves as input parameter to a sky (signal) rate model. Supplementary information from archived or quick-reduced spectra can be used to improve the ETC rate models and calibrations.}
\label{fig:etc-block}
\end{figure*}

Over the course of a science exposure, we will take Sky Monitor images through the r-band filter every 60 seconds during (lunar) bright time and every 200 seconds during dark and gray times. These cadences are set such that the Sky Monitor exposure time is $\sim \frac{1}{5}$ of the shortest science exposure time under these conditions. The fiber spots in the Sky Monitor images will be analyzed to measure the sky count and its error. Figure \ref{fig:etc-block} shows the software design of the ETC, which is integrated with the DESI ICS. 

In the event that a sky fiber needs to be repositioned to a blank region of the sky before the start of a science exposure, the ICS will first rotate the filter wheel to the position of the back illumination filter stack before turning on the back illumination LEDs. 

\section{COMMISSIONING AND OPERATION}
\label{sec:operation}
%2) Describe installation steps, particularly focusing. \\
%3) Describe results of on-sky data (David Kirkby's slides), i.e. we met image quality requirements for the ETC. \\
%4) Future steps? \\

Both units of the Sky Monitor have been installed at the Mayall Telescope, integrated with the DESI ICS, and have been in operation since late January 2020. The following describes the functional verification and performance results.

\subsection{Focusing the fiber tips}
The fiber block that secures the sky fibers is mounted on a manual XY translation stage. We manually adjusted the translation stage and determined the best focus of the fibers as seen through the r-band filter using ambient dome light. 

\begin{figure*}[h]
\centering
\includegraphics[width=0.8\textwidth]{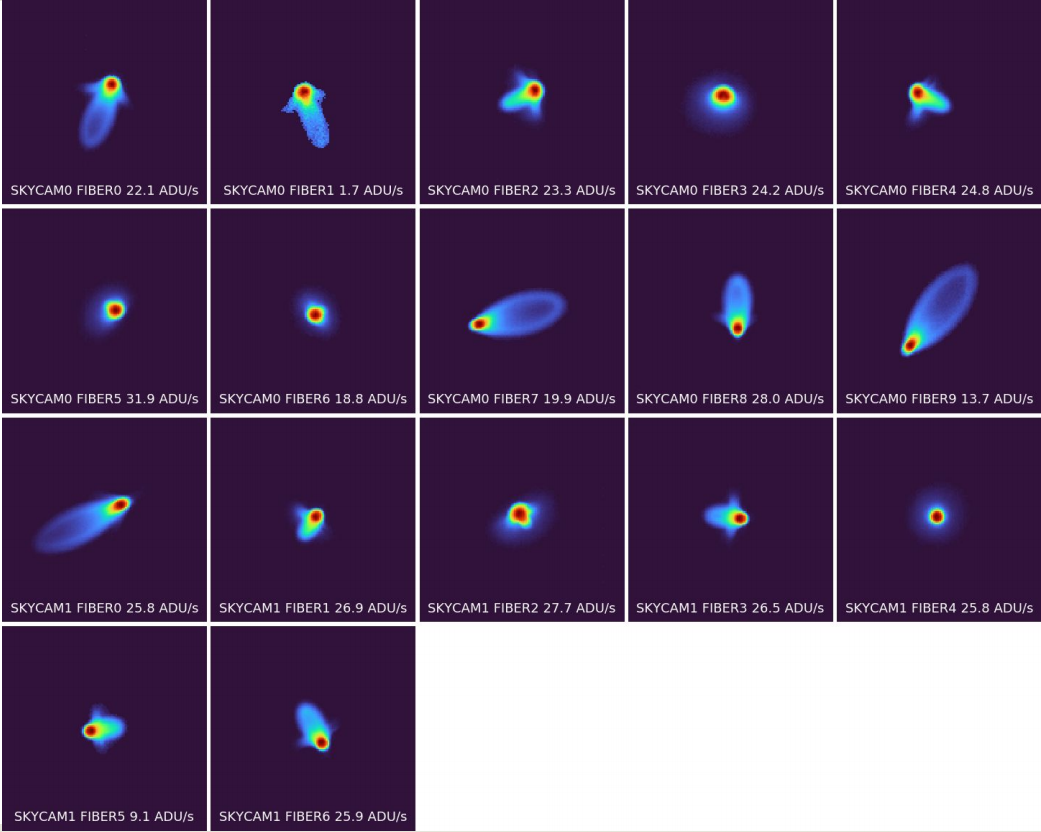}
\caption{\small Spot profile of the sky fibers at best focus position, taken during dark time at zenith position. With the exception of a few, the majority of the spots exhibit a fair amount of comatic aberration. This arises from off-axis placement and non-uniform depth of the fibers in the fiber block and imperfect tilt adjustment. Poor coupling of some fibers also results in a spread in the resulting count rates, as shown in the text of each thumbnail.}
\label{fig:spot-profile}
\end{figure*}

We discovered that the fiber tips appear to be at varying best focus positions and exhibit non-circular spot profiles due primarily to comatic aberrations. These arise due to a combination of off-axis placement of the fibers on the fiber block, imperfect tilt alignment (as our stage does not have tilt adjustment, we did it by eye), and non-uniform thread depth of the fibers in the fiber block. Therefore, we calculated the average best focus position and locked the system at this mean position. The spot aberrations are mitigated by measuring the spot profiles and the relative fiber throughput and accounting for them in the sky flux determination. Figure \ref{fig:spot-profile} shows the spot profiles of the seventeen sky fibers at the systemic best focus position. 

\subsection{Signal-to-noise ratio (SNR) of sky measurements}
The Sky Monitor is required to deliver sky images every 60s during (lunar) bright time and every 200s in dark and gray time. The sky level is expected to be measured with 4\% accuracy. Achieving better than 4\% will not be useful as the ETC is expected to be limited by calibration errors. Therefore, the SNR per fiber is required to be $>$ 10 so that the ETC can operate using a subset of the fibers, if necessary. 

Following the hardware installation of both Sky Monitor units in February 2020, the relevant software to operate it has also been successfully integrated with the DESI ICS. Since then, the Sky Monitor has been used to obtain commissioning data. 

We measured the performance of the Sky Monitor using the commissioning data collected under various observing conditions. Rather than the entire image of the Sky Monitor, the ETC only analyses smaller thumbnails centered on each fiber spot. The mean and background of each thumbnail image are first calculated to obtain an initial estimate of the variance per pixel, which is used to mask out hot pixels. After the fiber flux is fit, pixels with large chi-squares (due e.g. to cosmic rays) are removed. The fiber flux is then re-fit to obtain the final SNR per fiber. Finally, the weighted average flux and flux error are calculated to estimate the sky level. The data and analyses results are summarized in Table \ref{tab:data}, which demonstrate that the Sky Monitor delivers the required performance. 

\begin{table}[h]
\small
\caption{Sky Monitor data and results}\vspace{0.5ex}
\centering
\begin{tabular}{|c|c|c|c|c|c|c|}
\hline
Night      & \begin{tabular}[c]{@{}c@{}}Number of \\ exposures\end{tabular} & \begin{tabular}[c]{@{}c@{}}Exposure \\ time (sec)\end{tabular} & Conditions       & \begin{tabular}[c]{@{}c@{}}Moon \\ illumination\end{tabular} & \begin{tabular}[c]{@{}c@{}}Median \\ fiber SNR\end{tabular} & \begin{tabular}[c]{@{}c@{}}Median sky\\ level accuracy\end{tabular} \\ \hline
2020/01/26 & 10  & 60   & Zenith, dark     & 5\%    & 9.4   & 2.4\% \\ \hline
2020/02/26 & 10  & 60   & Moon set         & 11\%   & 11.3  & 2.0\% \\ \hline
2020/02/28 & 10  & 60   & Moon set, cloudy & 26\%   & 20.3  & 1.2\% \\ \hline
           & 10  & 60   & Dark, cloudy     & & 11.9 & 1.9\% \\ \hline
2020/02/29 & 10  & 60   & Moon up, cloudy  & 34\%   & 61.9  & 0.4\% \\ \hline
\end{tabular}
\label{tab:data}
\end{table}

\subsection{Back illumination}
\begin{figure*}[h]
\centering
\includegraphics[width=0.8\textwidth]{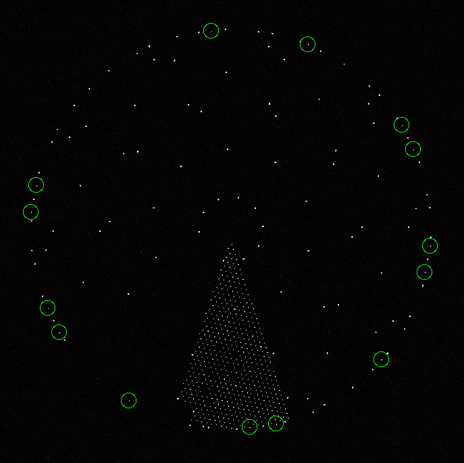}
\vspace{2mm}
\caption{\small The back-illuminated focal plane as imaged by the Fiber View Camera, consisting of a subset of the sky fibers that have been integrated with the Sky Monitor (green circles), one fully backlit petal (forming a pie-like structure), and all the fiducials.}
\label{fig:backillum}
\end{figure*}

We turned on the back illumination system and imaged the focal plane with the Fiber View Camera (FVC). The FVC relies on a spot detection code to accurately measure the center of each illuminated fiber and fiducial in its image. As such, sufficient and uniform counts over all the illuminated spots are essential. Figure \ref{fig:backillum} shows the back-illuminated focal plane with a subset of the backlit sky fibers, which are on average $\sim 2-3$ times brighter than the back-illuminated positioners and fiducials. 

%We plan to include a flashing capability to the current back  system to allow for brightness adjustment of the backlit sky fibers. We will perform this upgrade and test the intensity and uniformity of the sky fibers in a near future trip. 

\subsection{Light tightness}
Stray light should be minimized to allow for an accurate measurement of the faint sky flux. The Sky Monitor is located in the Large Coude Room, where all lights are turned off during normal night time operations. This, along with the enclosures, helps to reduce external light from leaking into the system. We fitted a baffle tube around the lens to remove internal stray light from the LED control electronics and status lights from the camera.

\begin{figure*}[h]
    \begin{minipage}[b]{0.50\linewidth}
        \centering
        \includegraphics[width=1.0\textwidth]{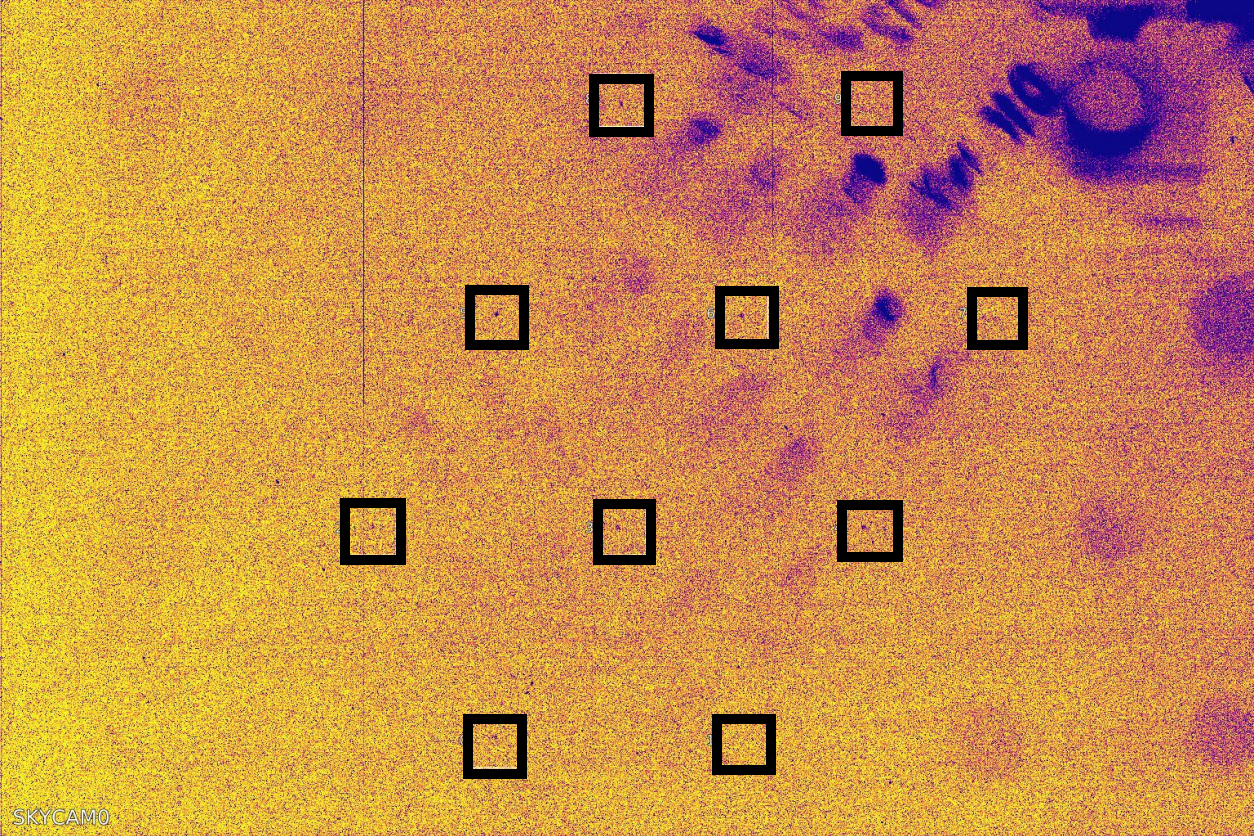}
    \end{minipage}
    \hspace{1mm}
    \begin{minipage}[b]{0.50\linewidth}
        \centering
        \includegraphics[width=1.0\textwidth]{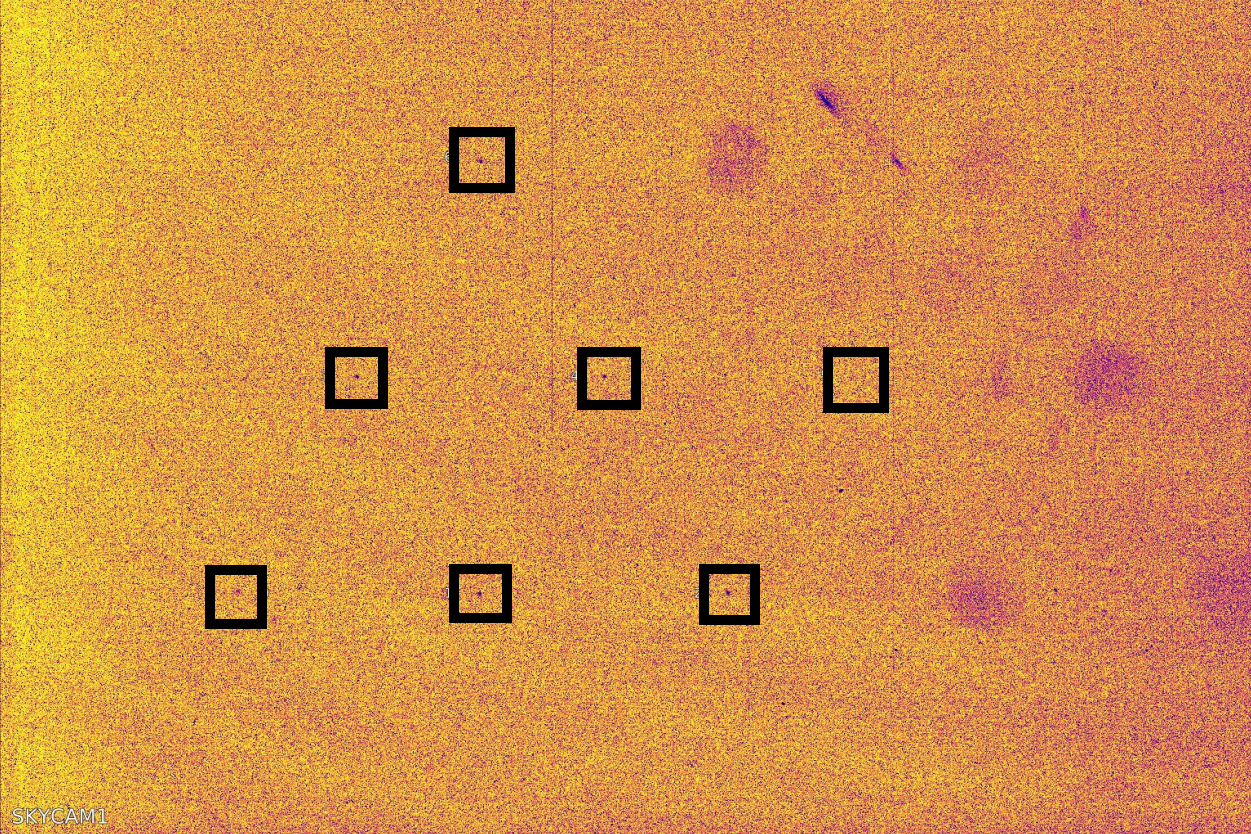}
    \end{minipage}
\vspace{2mm}
\caption{\small Raw images from both Sky Monitors show traces of stray light being reflected off shiny parts of the circuit board. These stray lights have minimal impact on the performance as they do not overlap with the analysis regions, denoted by the black squares, which are centered on the fiber spots. Note that dark areas represent regions with high flux and vice versa. }
\label{fig:straylight}
\end{figure*}

Figure \ref{fig:straylight} shows raw images from both Sky Monitors taken with 60 seconds exposures, where stray lights can be seen reflecting off shiny parts of the circuit board. However, they have minimal impact on the analyses as they do not land directly on any fiber spot.

\section{CONCLUSION}
\label{sec:conclusion}
The DESI Sky Monitor is an imaging system that tracks the night sky brightness as part of the dynamic ETC system. When combined with data from the DESI guide cameras, they provide a real-time SNR estimate for the observations. This allows the dynamic ETC to adjust the spectroscopic exposure time on-the-fly to produce data of uniform depth and to maximize observing efficiency. Both units have been successfully installed at the Mayall Telescope and integrated with the DESI ICS. We verified that all components are working appropriately and deliver the required performance necessary for the dynamic ETC.

To ensure the most optimum operation of the Sky Monitor, we plan to perform a number of future upgrades. One potential upgrade is to add a g-band filter (401 $-$ 550 nm) into the current setup. While an r-filter is selected as it covers the wavelength range of emission-line galaxies that are observed during dark time, the inclusion of a g-band filter would benefit low-redshift galaxies that are observed during bright time. Following this, we plan to configure the filters in the filter wheel to minimize the filter move time. As the filter wheel only rotates in one direction, one needs to rotate through empty slots in the filter wheel to land on a previous filter, where each move takes $\sim$ 2.5 seconds. Therefore, we plan to fill up the empty slots and configure the filters in a repeated ABAB (or ABCABC, if g-band filters are included) format. 

Finally, in designing the back illumination system, we used an appropriate number of LEDs so as to produce a brightness level comparable to the illumination level of the exposure shutter illuminator. The current system does not allow for brightness adjustment other than via the supplied voltage. To allow for flexibility, and considering that the fiber illuminator has flashing capability to accommodate a range of FVC exposure times, we consider adding an ability to turn the LEDs on and off at a certain duty cycle. All these future upgrades would entail extra commissioning steps, such as ensuring that the required performance level can be met with the new g-band filter and measuring the duty cycles required to achieve various back illumination levels. 

\acknowledgments % equivalent to \section*{ACKNOWLEDGMENTS} 
This research is supported by the Director, Office of Science, Office of High Energy Physics of the U.S. Department of Energy under Contract No. DE–AC02–05CH1123, and by the National Energy Research Scientific Computing Center, a DOE Office of Science User Facility under the same contract; additional support for DESI is provided by the U.S. National Science Foundation, Division of Astronomical Sciences under Contract No. AST-0950945 to the National Optical Astronomy Observatory; the Science and Technologies Facilities Council of the United Kingdom; the Gordon and Betty Moore Foundation; the Heising-Simons Foundation; the French Alternative Energies and Atomic Energy Commission (CEA); the National Council of Science and Technology of Mexico; the Ministry of Economy of Spain, and by the DESI Member Institutions. The authors are honored to be permitted to conduct astronomical research on Iolkam Du’ag (Kitt Peak), a mountain with particular significance to the Tohono O’odham Nation.  

% References
\bibliography{main} % bibliography data in report.bib
\bibliographystyle{spiebib} % makes bibtex use spiebib.bst

\end{document}